\documentclass[superscriptaddress,letterpaper,twoside,twoclumn,showpacs,amsmath,amssymb,aps,prl,floatfix,showkeys,reprint,]{revtex4-2}
\usepackage{graphicx}% Include figure files
\usepackage{dcolumn}% Align table columns on decimal point
\usepackage{bm}% bold math
\usepackage{amsmath}
\usepackage{amssymb}
\usepackage{multirow}
\usepackage{color}
\usepackage{textcmds} % allows \qq{}
\usepackage{booktabs}
\usepackage{tabularx}
\usepackage[english]{babel}
\usepackage[utf8]{inputenc}
\usepackage{siunitx}

\usepackage[normalem]{ulem}
\usepackage{xcolor}

\newcommand{\ie}{i.\,e.,\,}
\newcommand{\eg}{e.\,g.,\,}
\newcommand{\fig}{Fig.\,}

\begin{document}
\title{Zero-field skyrmionic states and in-field edge-skyrmions induced by boundary tuning}

\author{Jonas Spethmann}
\email{jspethma@physnet.uni-hamburg.de}
\affiliation{Department of Physics, University of Hamburg, 20355 Hamburg, Germany}
\author{Elena Y. Vedmedenko}
\affiliation{Department of Physics, University of Hamburg, 20355 Hamburg, Germany}
\author{Roland Wiesendanger}
\affiliation{Department of Physics, University of Hamburg, 20355 Hamburg, Germany}
\author{Andr\'e Kubetzka}
%\email{kubetzka@physnet.uni-hamburg.de}
\affiliation{Department of Physics, University of Hamburg, 20355 Hamburg, Germany}
\author{Kirsten von Bergmann}
%\email{kbergman@physnet.uni-hamburg.de}
\affiliation{Department of Physics, University of Hamburg, 20355 Hamburg, Germany}
	   
		%\email[Corresp.\ author: ]

\date{\today}

\begin{abstract} 
When magnetic skyrmions are moved via currents, they do not strictly travel along the path of the current, instead their motion also gains a transverse component. This so-called skyrmion Hall effect can be detrimental in potential skyrmion devices because it drives skyrmions towards the edge of their hosting material where they face potential annihilation. Here we experimentally modify a skyrmion model system|an atomic Pd/Fe bilayer on Ir(111)|by decorating the film edge with ferromagnetic Co/Fe patches. Employing spin-polarized scanning tunneling microscopy, we demonstrate that this ferromagnetic rim prevents skyrmion annihilation at the film edge and stabilizes skyrmions and target states in zero field. Furthermore, in an external magnetic field the Co/Fe rim can give rise to skyrmions pinned to the film edge. Spin dynamics simulations reveal how a combination of different attractive and repulsive skyrmion-edge interactions can induce such an edge-pinning effect for skyrmions.

 % 148/150 words
\end{abstract}
\maketitle

\section{Introduction}

Magnetic skyrmions are whirling spin textures with topological properties that make them robust against external perturbations. They were first described theoretically~\cite{bogdanov1989,bogdanov1994} and later found experimentally in systems with broken structural inversion symmetry such as chiral magnets~\cite{muhlbauerS2009,yuN2010} or magnetic ultra-thin films~\cite{heinzeNP2011,rommingS2013a}. Their well investigated transport properties~\cite{sampaioNN2013,nagaosaNN2013,kangIEEE2016,everschorJAP2018,zhangIOP2020} make them interesting candidates as information carriers in future generation spintronic devices, such as racetrack memories~\cite{parkinS2008,fertNN2013}, logic gates~\cite{zhangSR2015,luoNL2018,songIEEE2021} or recently proposed applications that exploit Brownian skyrmion motion~\cite{pinnaPRA2018,zazvorkaNN2019,jibikiAPL2020}. Skyrmions are typically stabilized by an interplay of exchange interaction, Dzyaloshinskii-Moriya interaction (DMI)~\cite{dzyaloshinskii1957,moriya1960} and perpendicular magnetocrystalline anisotropy (PMA). Frustrated exchange interactions can further enhance the skyrmion stability~\cite{malottkiSR2017,yuanPRB2017} and even allow their formation in zero magnetic field~\cite{meyerNC2019}. The size of skyrmions ranges from sub-10~nm in epitaxially grown ultra-thin metal films, in which skyrmions are stable at cryogenic temperatures~\cite{heinzeNP2011,rommingS2013a,hsuNN2017,meyerNC2019}, to room-temperature skyrmions in multilayer films with diameters between 30~nm and 2~$\mu$m~\cite{jiangS2015,wooNM2016,soumyanarayananNM2017,oharaNL2021}. 

One limitation that potential skyrmion applications may face is the occurrence of the skyrmion Hall effect (SkHE), which arises due to the topological nature of the skyrmions and drives them towards the edge of their hosting material when they are moved via currents~\cite{schulzNP2012,jiangNP2017,litziusNP2017}. This can cause skyrmion annihilation at the edge~\cite{iwasakiNN2013,purnamaSR2015,bessarabSR2018}, which would be detrimental to their use in such devices. One proposal to mitigate this problem is to use skyrmionic states with compensated topological charge ($Q=0$) that theoretically do not show the SkHE, \eg antiferromagnetic skyrmions~\cite{barkerPRL2016,zhangNC2016, dohiNC2019} or $2\pi$ skyrmions,  which are also called target states~\cite{zhengPRL2017,kolesnikovSR2018,hagemeisterPRB2018,cortes-ortunoPRB2019a}. Alternatively it was proposed to manipulate the magnetic properties of the skyrmion material itself by creating a potential well that guides the skyrmion along a desired pathway~\cite{purnamaSR2015, laiSR2017,toscano2020}. Recently two groups have made experimental advances in this regard using sputter-deposited thin film systems: One group created skyrmion-stabilizing tracks with weakened PMA and DMI by slightly altering their Pt/Co/MgO film with focused He+ ion beam irradation~\cite{jugeNL2021}. The other group was able to grow a multilayer film patterned with high-PMA areas that repel skyrmions~\cite{oharaNL2021}.

\begin{figure}[tb] 
	\centering
	\includegraphics[width=0.45\textwidth]{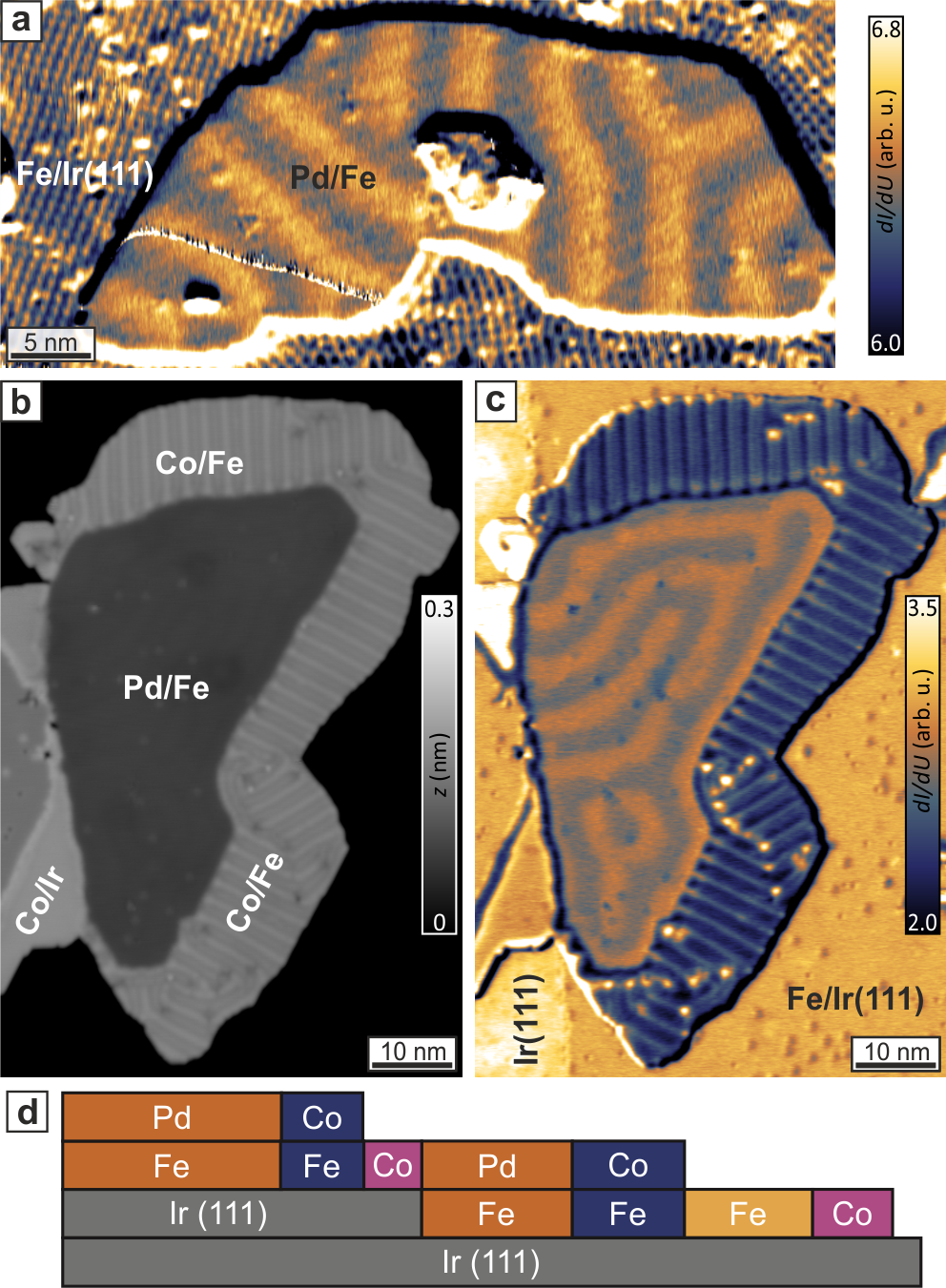}
	\caption{The Pd/Fe/Ir(111) system with Co/Fe-decorated edges. \textbf{a}~$dI/dU$ map of a Pd/Fe island on Ir(111) showing the spin spiral state ($I=2$~nA, $U=15$~mV). \textbf{b,~c}~constant-current image and simultaneously recorded $dI/dU$ map of a Pd/Fe island surrounded by Co/Fe ($I=2$~nA, $U=-310$~mV); with the Co/Fe rim the spin spiral stripes are oriented parallel to the island edges. \textbf{d}~Sketch of the sample composition.}
	\label{fig1}
\end{figure}

Here we investigate skyrmion-edge interactions in an epitaxially grown model-type system using spin-polarized scanning tunneling microscopy (SP-STM), which employs the tunneling magnetoresistance effect (TMR). Our model system is an atomic Pd/Fe bilayer on the surface of an Ir(111) single crystal, which|at cryogenic temperatures|has a cycloidal spin spiral as its magnetic ground state. Under application of an external magnetic field the spin spiral turns first into a skyrmion lattice and then into individual skyrmions with diameters of 2-3~nm~\cite{rommingS2013a, rommingPRL2015a}. Figure \ref{fig1}a shows an SP-STM image of such a Pd/Fe island. The spin spiral can be identified by periodically occurring stripes of darker and brighter signal. The island shape has a strong effect on the propagation direction of the spin spiral: The spiral stripes are oriented along the shorter island extension and bend in such a way, that they connect perpendicular to the Pd island edge. This configuration is preferred over a parallel alignment to the island edge, because it avoids the formation of a ferromagnetically ordered island rim~\cite{schmidtNJP2016}, and in this way allows the DMI energy to be further reduced by edge tilting effects~\cite{rohartPRB2013a, Hagemeister2016}. 

These results show that the island edge has a clear influence onto the details of the magnetic state and the question arises, whether the island edge can be utilized to manipulate the magnetic state inside the Pd/Fe island in a desired fashion.
In this work we tune the properties of the Pd/Fe edge by growing a self-assembled ferromagnetic Co/Fe bilayer adjacent to it. We find that the Co/Fe bilayer has an immediate effect on the spin spiral ground state, stabilizes skyrmions and target states at zero field and gives rise to skyrmions pinned to the Pd/Fe island edge in applied magnetic fields. Finally we perform spin dynamics simulations to investigate the role of different magnetic parameters in causing these edge effects.

\section{Results}

\subsection{Tuning the Pd/Fe island edge.}  

We attempt a tuning of the Pd/Fe island edges by decorating them with a Co/Fe bilayer. For this, approximately 0.2 atomic layers of Co are deposited onto an existing Pd/Fe/Ir(111) sample at room temperature. The Co grows either on the uncovered Ir surface or on top of the Fe monolayer, where it preferentially attaches itself to the Pd island edge, see \fig\ref{fig1}d. In this way free standing Pd/Fe islands are typically fully surrounded by Co/Fe; only rarely do we observe Co islands on top of Pd/Fe. Figure \ref{fig1}b shows an STM constant-current image of a Pd/Fe island surrounded by Co/Fe at the right and upper edge, while the left island rim is adjacent to a buried Ir(111) step edge.

Note that the Co/Fe areas show regular lines that originate from a structural reconstruction: While the Co monolayer on Ir(111) is pseudomorphic, the Co/Fe bilayer releases lattice strain by forming dislocation lines perpendicular to the close-packed atomic rows. The uniaxial reconstruction pattern has a period of 2.5~nm and is present in three rotational domains due to the hexagonal symmetry of the surface. We find that the Co/Fe bilayer is ferromagnetic, with two contrast levels per rotational domain and up to six different contrast levels in total, see Supplementary \fig S1. We conclude that the easy magnetization axis of Co/Fe must be in-plane and from measurements with a known tip magnetization axis we derive the easy magnetization axis of Co/Fe to be perpendicular to its dislocation lines, see Supplementary \fig S2. 

Figure \ref{fig1}c shows a differential conductance ($dI/dU$) map of the same Pd/Fe island as shown in b. Here the magnetic contrast of the Pd/Fe spin spiral in the magnetic virgin state is observed. Comparison with the island in \fig\ref{fig1}a demonstrates that the adjacent Co/Fe strongly affects the path of the spin spiral stripes, which are now oriented parallel to the Co/Fe-decorated island edges. Only at the left island edge, which is not next to a Co/Fe area, the spiral stripes seem to prefer an orthogonal orientation. Due to the shape of the island the spin spiral in the interior of the island is also affected by the Co/Fe edge and is much more disordered when compared to \fig\ref{fig1}a. 

\begin{figure}[tbh] 
	\centering
	\includegraphics[width=0.45\textwidth]{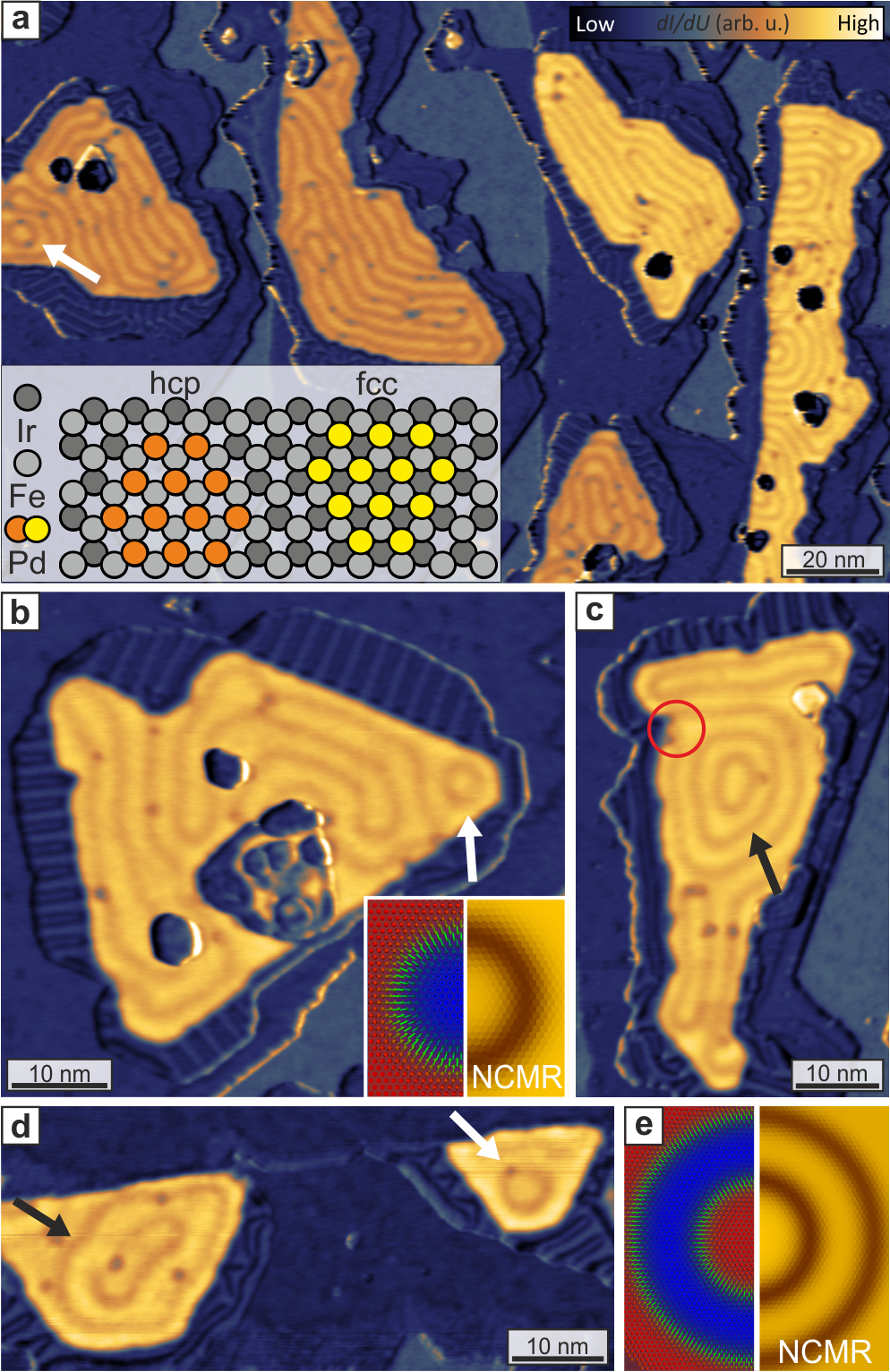}
	\caption{Skyrmions and target states in the magnetic virgin state of Co/Fe-decorated Pd/Fe/Ir(111) ($dI/dU$ maps: $I=3$~nA, $U=710$~mV). \textbf{a} Spin spirals in hcp-stacked (orange) and fcc-stacked (yellow) Pd/Fe islands are affected by the Co/Fe in a similar fashion; inset shows a sketch of the two possible stacking orders. \textbf{b}~Pd/Fe island with a zero-field skyrmion in a corner; inset is split in the magnetic structure of a skyrmion (left) and expected NCMR contrast (right). \textbf{c}~Magnetic target state, almost fully encircled by a third ring. \textbf{d}~Target state and zero-field skyrmion confined inside relatively small Pd/Fe islands. \textbf{e}~Image of a magnetic target state split in half to show the spin structure (left) and corresponding NCMR contrast (right).}
	\label{fig2}
\end{figure}
%$dI/dU$ colour scales: (a)~$0.6-2.3$, (b)~$1.3-6.8$, (c)~$1.3-9.3$, (d)~$1-6.5$ (a.u.)

\begin{figure*}[tbh] 
	\centering
	\includegraphics[width=1\textwidth]{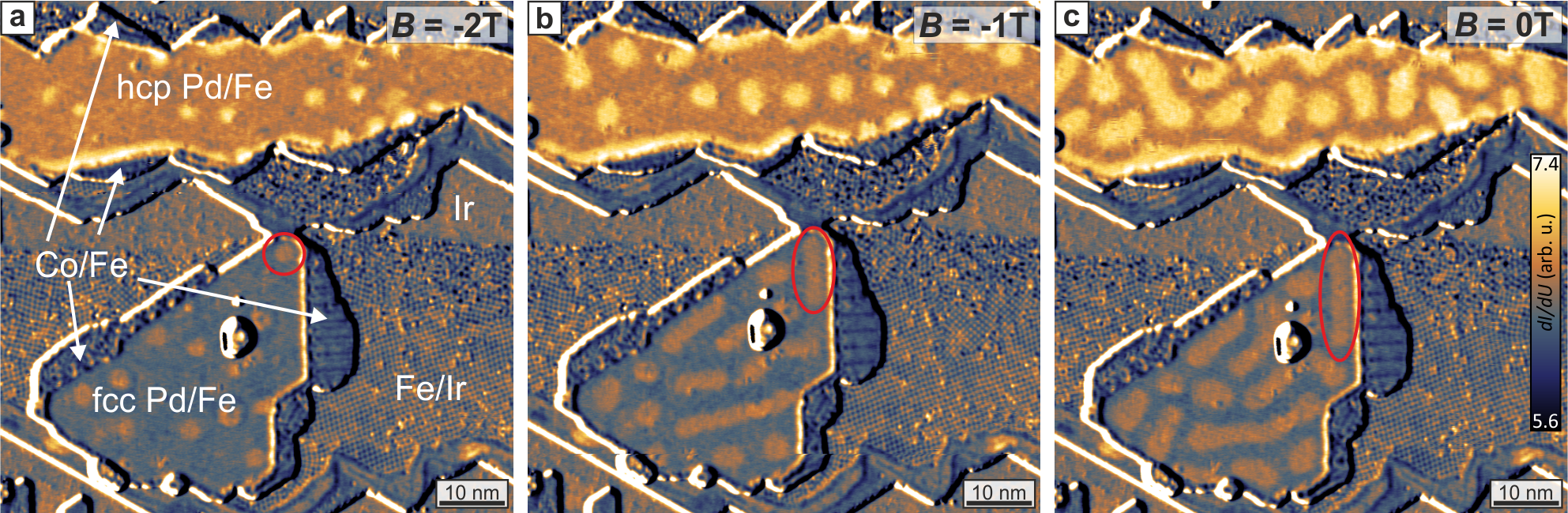}
	\caption{Remanent skyrmions after magnetic field-sweep ($dI/dU$ maps: $I=3$~nA, $U=20$~mV). \textbf{a}~Two differently stacked Pd/Fe islands surrounded by Co/Fe in an external field of $B=-2$~T; skyrmions can be written or deleted by scanning the sample surface with a sufficiently high bias voltage~\cite{rommingS2013a}. \textbf{b}~The same two islands after reducing the external field to $B=-1$~T. \textbf{c}~Final image at $B=0$~T; Many skyrmions have survived and are now in a metastable state. A skyrmion performing a partial strip-out parallel to the island edge is encircled in red.}
	\label{fig3}
\end{figure*}

\subsection{Zero-field skyrmions and target states.} In \fig\ref{fig2}a an overview image with several Pd/Fe islands|each surrounded by Co/Fe|is shown. At the bias voltage chosen here, one can also discriminate between hcp-stacked (orange) and fcc-stacked (yellow) Pd/Fe islands, see inset. These two Pd/Fe stackings exhibit subtle differences in their magnetic interaction parameters which lead to, \eg a slightly larger spiral period in the case of hcp-stacked Pd/Fe. Note that these images were measured with a non-magnetic tip, utilizing the non-collinear magnetoresistance effect (NCMR), which is caused by a mixing of spin channels that occurs in non-collinear spin structures and is highly sensitive to fast-rotating spin textures~\cite{hannekenNN2015a}. The contrast originating from the NCMR effect differs from that of the TMR: Due to the PMA of Pd/Fe/Ir(111), the spin spirals in the system are inhomogeneous, \ie the out-of-plane sections are more collinear, while the nearest-neighbor angle in in-plane areas becomes larger. When imaged with NCMR contrast, the tip is therefore equally sensitive to all in-plane sections of the spiral, which effectively halves the measured distance between the dark stripes compared to TMR measurements. As shown in the inset of \fig\ref{fig2}b, skyrmions can also be detected using the NCMR effect: Depending on their size they can appear as dark rings or dots, because again the in-plane sections of a skyrmion are strongly non-collinear while the skyrmion center is mostly collinear below a critical external field~\cite{hannekenNN2015a}.

In both Pd/Fe stackings, the Co/Fe-decorated edges appear to affect the spin spirals in a similar fashion: The spin spiral stripes prefer to be parallel to the island edges, which often leads to the formation of more exotic spin structures in the magnetic virgin state of the system, see \fig\ref{fig2}a-d, including skyrmions (white arrows) and target states (black arrows). The appearance of skyrmions in the virgin state is remarkable because typically external fields above 1~T are necessary to stabilize skyrmions in Pd/Fe/Ir(111). These zero-field skyrmions are often observed in rather small Pd/Fe islands or corners of larger islands, where the surrounding Co/Fe seemingly prevents the skyrmion from connecting to the island edge. Zero-field target states|see \fig\ref{fig2}e for the spin structure and expected NCMR contrast|have not been observed in Pd/Fe/Ir(111) prior to this publication, despite an effort to stabilize them in confined geometries \cite{cortes-ortunoPRB2019a}. However, in samples with added Co/Fe at the boundaries, they appear frequently in zero field. Depending on the Pd/Fe island size and geometry, the target states often deviate from their ideal circular shape and their diameter may vary in the range of 10-20~nm. The target state shown in \fig\ref{fig2}c is surrounded by an almost complete third ring, which would make it a $3\pi$ skyrmion, were it not for the small section that is connected to the island rim (see red circle). Nevertheless, these findings suggest that it may be possible to stabilize and study $k\pi$ skyrmions with $k>2$ by carefully adjusting the size and shape of the Pd/Fe islands surrounded by Co/Fe.

Before concluding this part about zero-field measurements, we want to emphasize that we observe a strong impact of Co/Fe onto the magnetic configuration within Pd/Fe. The presence of a ferromagnetic domain at the island edge, seems to impose the formation of a ferromagnetically ordered rim in Pd/Fe, which forces the spin spiral to change its propagation direction at the edge and also helps to stabilize different skyrmionic states in zero-field.

\begin{figure}[tbh] 
	\centering
	\includegraphics[width=0.45\textwidth]{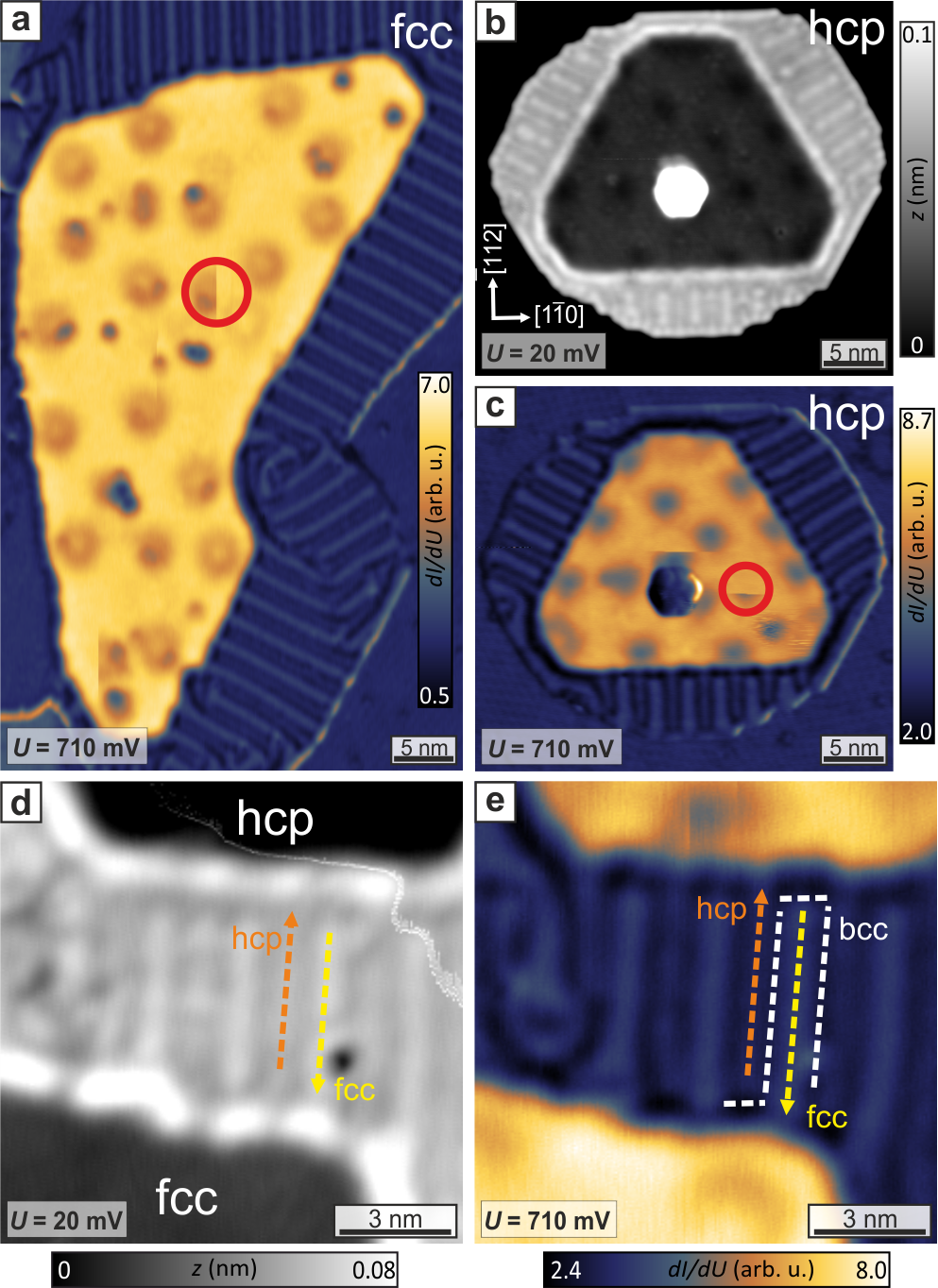}
	\caption{Differences between fcc- an hcp-stacked Pd/Fe islands and details of the Co/Fe transition area ($I=3$~nA, $B=2$~T). \textbf{a}~$dI/dU$ map of an fcc-stacked Pd/Fe island with dominant NCMR contrast in an applied external field. \textbf{b}~Constant-current SP-STM image of a triangular hcp-stacked Pd/Fe island, surrounded by Co/Fe. \textbf{c}~$dI/dU$ map of the same island, scanned at a different bias voltage. \textbf{d}~Close-up constant-current image of a Co/Fe area located between a hcp-stacked (top) and fcc-stacked (bottom) Pd/Fe island; the reconstruction pattern changes close to the island rim. \textbf{e}~$dI/dU$ map of the same area, the different hollow site sections and the bcc(110)-like bridge sites can be clearly distinguished.}
	\label{fig4}
\end{figure}

\begin{figure}[tbh] 
	\centering
	\includegraphics[width=0.45\textwidth]{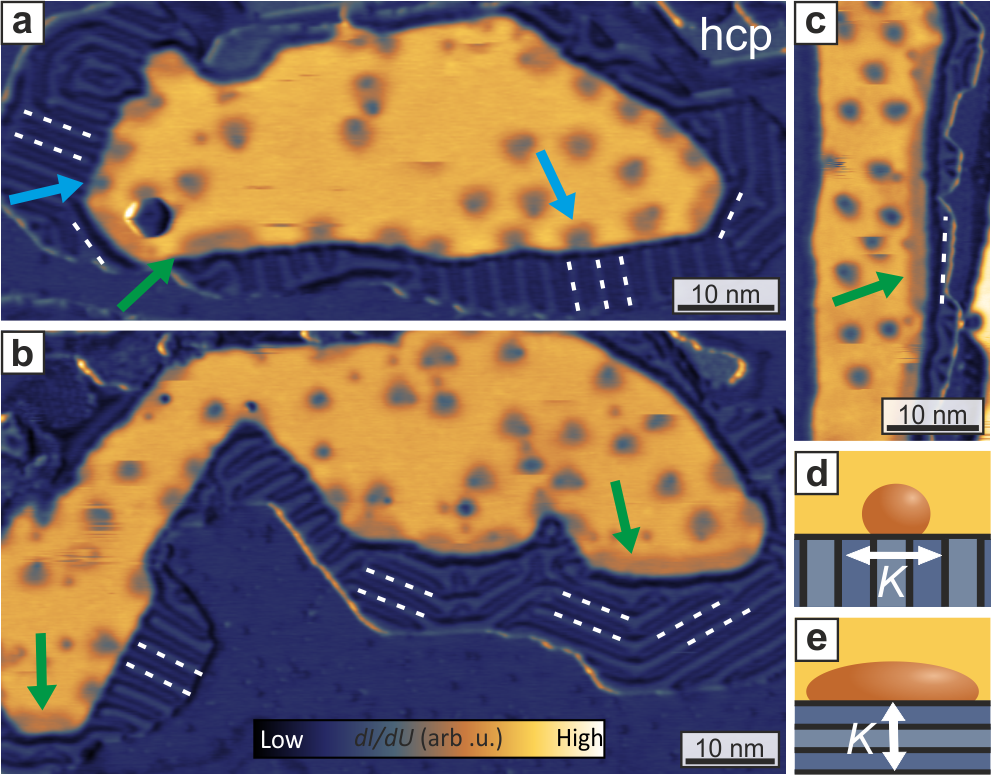}
	\caption{Details of the skyrmion-edge interaction in Co/Fe-decorated hcp-stacked Pd/Fe. \textbf{a-c}~$dI/dU$ maps (NCMR) of hcp Pd/Fe islands with localized (blue arrows) and more striped-out (green arrows) skyrmion-like objects at the island edge. Dashed lines indicate the direction of the dislocation lines in Co/Fe ($I=3$~nA, $U=710$~mV, $B=2$~T). \textbf{d,e}~simplified sketches showing localized- and striped-out skyrmions pinned at two different Co/Fe edge types.}
	\label{fig5}
\end{figure}

\subsection{Skyrmions at the edge.} Whereas we have seen that the Co/Fe rim stabilizes skyrmions in zero-field, in the following we want to study the impact of the ferromagnetic rim on field-induced skyrmions. Figure \ref{fig3}a shows a measurement of two differently stacked Pd/Fe islands in a perpendicular magnetic field of $B=-2$~T. Both islands are surrounded by thin patches of Co/Fe along most parts of the island boundary. Instead of spin spirals, the Pd/Fe islands are now filled with several skyrmions in an otherwise ferromagnetic background. When the external field is reduced to $-1$~T (b) and finally $0$~T (c), the skyrmion size increases and some skyrmions perform a partial strip-out (typically parallel to the island edge). However, even at $0$~T many skyrmions survive, likely because they cannot connect to the island edge, which is the dominant pathway of skyrmion annihilation in low magnetic fields~\cite{bessarabSR2018}. Such a stabilizing effect was also predicted for the edge tilt occurring at the boundary of magnetic materials due to DMI~\cite{rohartPRB2013a,cortes-ortunoPRB2019a}. Yet, experimentally we observe that in free-standing Pd/Fe islands, skyrmions typically strip out and connect to the island rim upon a decrease of $B$, forming spiral stripes perpendicular to the edge, see Supplementary \fig S3. We conclude that skyrmion escape via the island edge is effectively inhibited by the adjacent ferromagnetic Co/Fe. 

In both Pd/Fe islands shown in \fig\ref{fig3} the skyrmions survive in remanece at $0$~T due to the modification at the island rim, yet we find significant differences in the skyrmion-edge interaction between the two Pd/Fe stacking types. In fcc Pd/Fe the magnetic texture is typically separated from the rim and the skyrmions keep a distance of at least 1-2~nm to the island boundary. In hcp Pd/Fe, on the other hand, the rim is populated with skyrmion-like spin textures at $B=2$~T. When the field is lowered, these objects eventually strip out further and form more continuous single spiral-like stripes, that seemingly prevent skyrmions from inside the island to connect to the island edge. Even in measurements using a relatively high bias voltage of $U$~=~710~mV, see \fig\ref{fig4}a,c, which has been shown to affect the magnetic structure and can induce skyrmion displacement, nucleation or annihilation (red circles), the edge-skyrmions at the Co/Fe rim remain relatively stable. 

In order to investigate the origin of these different skyrmion-edge interactions in hcp- and fcc-stacked Pd/Fe, we will now look more closely at the structure of the Co/Fe reconstruction and how it connects to the Pd/Fe island edge. Figure \ref{fig4}b and c show a constant-current image and a $dI/dU$ map of the same hcp-stacked Pd island surrounded by three rotational domains of the Co/Fe reconstruction, while \fig\ref{fig4}d and e show a zoom-in of a Co/Fe area located between a hcp-stacked and fcc-stacked Pd/Fe island. The contrast on the Co/Fe reconstruction shows three distinct sections, reminiscent of the reconstructed Fe double layer on Ir(111)~\cite{HsuPRL2016b}. Here we have developed a similar model for the Co/Fe reconstruction, where the Co atoms continuously shift between hcp and fcc hollow sites separated by bcc(110)-like bridge sections in between with a lattice compression of 10\%, see Supplementary \fig S4. 

The reconstructed Co/Fe does not directly attach itself to the Pd/Fe island edge, instead we observe a narrow transition region. This can be seen \eg in \fig\ref{fig4}b, which shows an approximately 1~nm wide bright rim surrounding the Pd/Fe island. The bias dependence of the recorded contrast, see also Supplementary \fig S5, makes it difficult to interpret such STM data, yet we propose that the first couple of Co atomic rows at the Pd/Fe island edge grow pseudomorphic|with the same stacking as the adjacent Pd/Fe|and that only further from the edge the Co/Fe transitions into its fully reconstructed phase. On both sides of the Co/Fe stripe in \fig \ref{fig4}d,e different sections of the reconstructed Co/Fe merge with the respective pseudomorphic Co/Fe transition region. This leads to the displayed assignment of hcp and fcc lines in the reconstructed Co/Fe. We expect that different magnetic properties of the differently stacked pseudomorphic Co/Fe areas are fundamental to understand why skyrmions pin to the island edge or not, as observed for hcp- and fcc-stacked Pd/Fe, respectively. 

Further examples of hcp-stacked Pd/Fe islands, see \fig\ref{fig5}a-c, show that in an applied field the island rim is actually populated with different skyrmionic objects. On the one hand there are localized edge-skyrmions (blue arrows) that have a similar size and shape as the skyrmions in the island interior, on the other hand we also find more striped-out skyrmion-like objects (green arrows) that resemble a single spin spiral stripe. The localized and striped-out skyrmions typically pin to different parts of the Co/Fe rim. Through analysis of all available data, we find a general trend: In areas where the dislocation lines in the adjacent Co/Fe (see dashed lines) are perpendicular to the Pd/Fe island edge, we typically find localized skyrmions. On the other hand, if the dislocation lines run parallel or with an angle of up to $30^\circ$ towards the rim, we dominantly find more striped-out skyrmions. \fig\ref{fig5}d and e schematically illustrate this trend, and also display the relation to the easy magnetization axis of Co/Fe. These findings are remarkable, because they indicate that it is possible to realize several different magnetic states at the island rim by slight modification of the surrounding material properties.

\subsection{Skyrmion-edge interactions.} The findings of the previous section have inspired us to investigate the role of different magnetic interactions with regards to their ability to pin or repel skyrmions at the island edge. We employ atomistic spin dynamics simulations~\cite{montecrystal} using the following standard Hamiltonian for skyrmions in ultra-thin films~\cite{hagemeisterPRB2018}:

\begin{equation}
	\begin{split}
		H = -J\displaystyle\sum_{\langle i,j \rangle}(\mathbf{S}_i \cdot \mathbf{S}_j) -\displaystyle\sum_{\langle i,j \rangle}\mathbf{D}_{i,j}\cdot(\mathbf{S}_i \times \mathbf{S}_j) \\ -\displaystyle\sum_{i}K_z\cdot S^2_{i,z} -\displaystyle\sum_{i}K_x\cdot S^2_{i,x} -\mu\displaystyle\sum_{i}(\mathbf{B} \cdot \mathbf{S}_i) 
	\end{split}
\end{equation}

\noindent where $S_i$ and $S_j$ are classcial spins with $|S|=1$, $J$ is an effective Heisenberg exchange interaction coefficient mapped on the nearest-neighbor exchange, $\mathbf{D}_{i,j}$ is the Dzyaloshinksii-Moriya vector acting on nearest neighbour spins, $K_z$ is the uniaxial anisotropy parameter and $\mu$ is the strength of the magnetic moment. The film system consisting of Pd/Fe and Co/Fe is mapped onto a single hexagonal monolayer with three different regions: A Pd/Fe area which hosts spin spirals and skyrmions, a ferromagnetic Co/Fe area with a uniaxial in-plane anisotropy $K_x$ and lastly a narrow pseudomorphic transition region in between to represent the changes of the Co/Fe structure close to the Pd/Fe island. We use a magnetic moment of $\mu=3~\mu_\mathrm{B}$ as calculated for Pd/Fe~\cite{dupeNC2014a} in the entire layer and vary the magnetic interaction parameters to mimic the different materials, see Tab.\,\ref{table1}. For Pd/Fe we use experimentally determined magnetic parameters~\cite{rommingPRL2015a}. The magnetic parameters of the reconstructed Co/Fe are unknown, and we chose values that reflect our experimental findings. In particular we find that Co/Fe is nearly aligned with an out-of-plane applied magnetic field of 2T, see methods and Supplementary \fig S6. The pseudomorphic Co/Fe area in between Pd/Fe and reconstructed Co/Fe is represented by 4 atomic rows. Whereas experimentally it is difficult to precisely determine the width of this transition area, our simulations show that the exact number of atomic rows is not crucial for the results presented in the following. Because this Co/Fe area is not reconstructed an isotropic easy plane anisotropy of $K_z=-0.25$~meV/atom is chosen.

\begin{figure}[b] 
	\centering
	\includegraphics[width=0.45\textwidth]{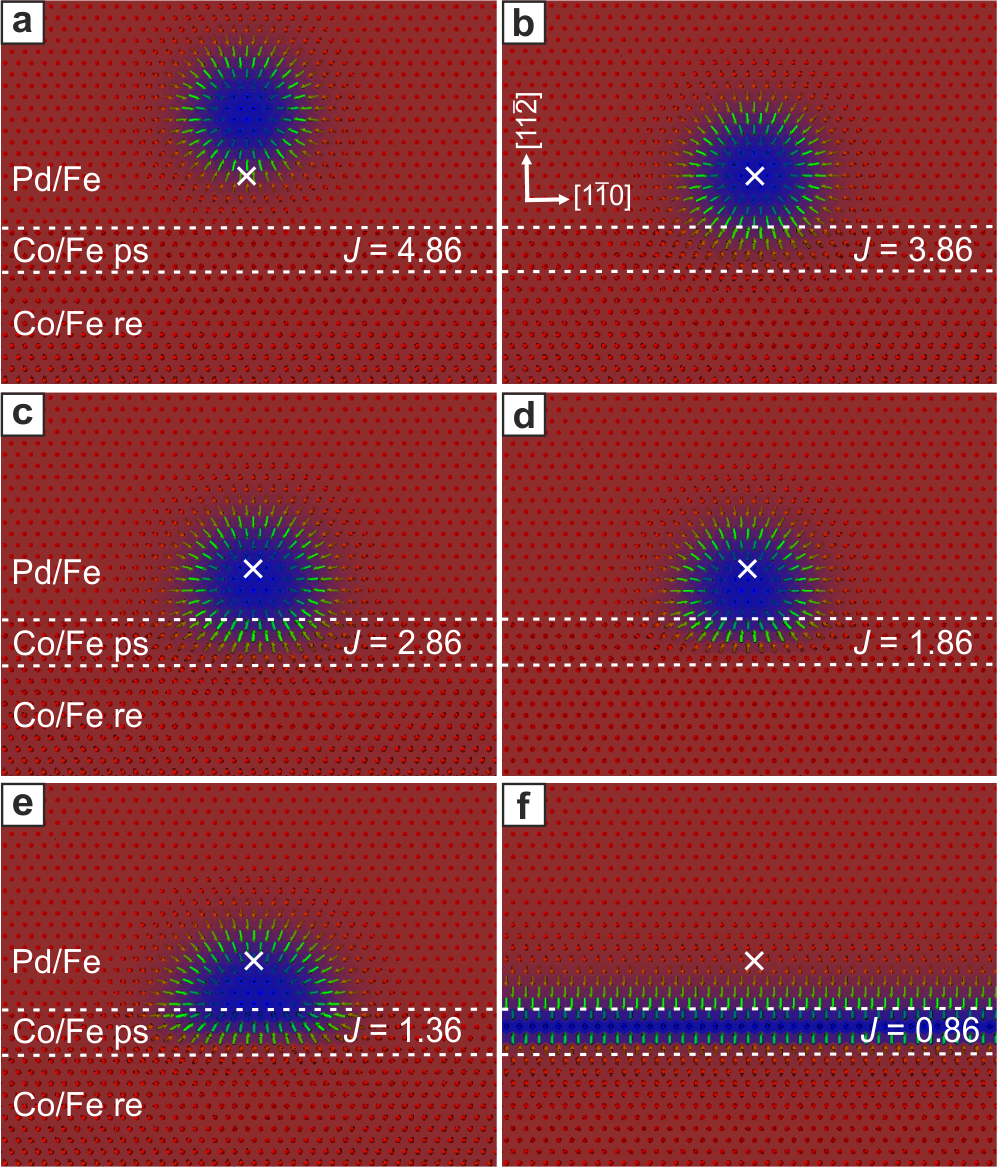}
	\caption{Spin dynamics simulations with different strengths for the exchange interaction $J$ in the pseudomorphic Co/Fe area. Red and blue dots represent spins pointing in opposite out-of-plane directions. In-plane pointing spins are visualized as green arrows. The hexagonal lattice is split in three distinct sections, as marked by the dashed lines, with different magnetic interaction parameters, see Tab.\,\ref{table1}. Initially, the skyrmion was written at the position of the white cross. All simulations were performed at $0$~K and $B=2$~T. The displayed area was cut from a larger simulation area, see methods.}
	\label{fig6}
\end{figure}

\begin{table}[t] 
	\vspace{-3.5mm}
	\centering
	\caption {Magnetic interaction parameters of the three different areas of the simulated layer in \fig\ref{fig6} (meV/atom).}
	\begin{ruledtabular}
		\begin{tabular}{l|l|l|l} 
			Bilayer & Exchange & DMI & Anisotropy\\
			\hline
			Pd/Fe & $J=2.86$ & $D=0.76$ & $K_z=0.4$, $K_x=0$\\
			Co/Fe ps & $J=$varied & $D=0.76$ & $K_z=-0.25$, $K_x=0$\\
			Co/Fe re & $J=7.86$ & $D=0.76$ & $K_z=0.4$, $K_x=0.65$\\
		\end{tabular}
	\end{ruledtabular}
	\label{table1}
\end{table}

\begin{figure*}[tbh] 
	\centering	
	\includegraphics[width=1\textwidth]{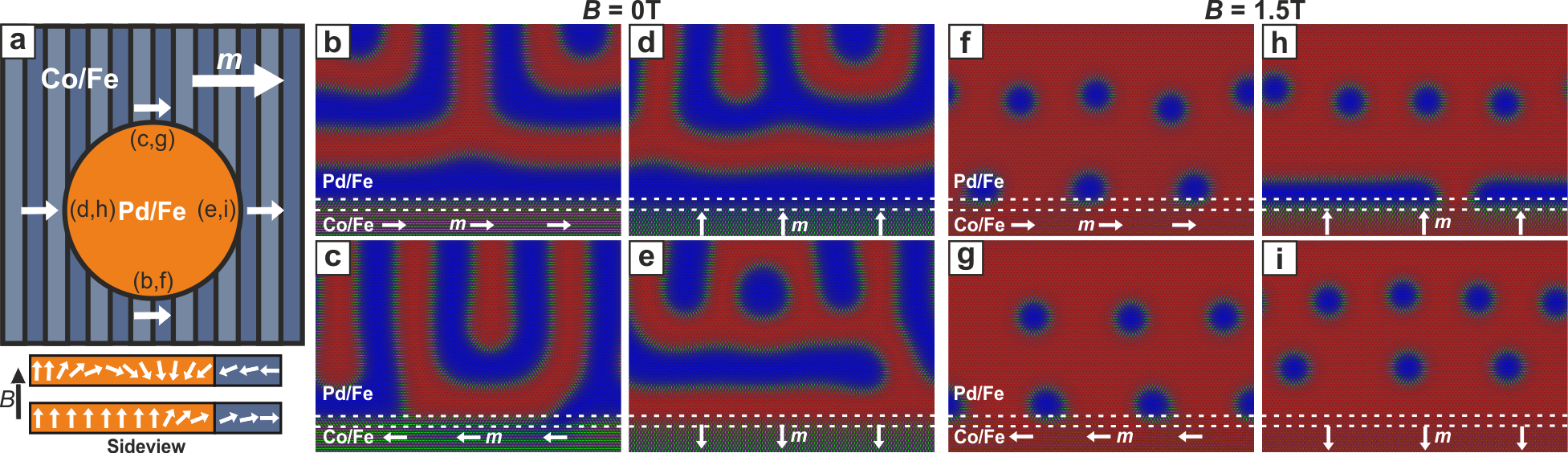}
	\caption{Rotating the in-plane magnetization component (white arrows) in the reconstructed Co/Fe. \textbf{a} Schematic overview of the four magnetization directions (top) and clockwise rotation at the edge with and without incorporating an edge-skyrmion (bottom). \textbf{b-e} Simulations at zero field and \textbf{f-i} at $B=1.5$~T. The hexagonal lattice has periodic boundary conditions in the horizontal direction and is split in three distinct areas with the magnetic parameters as shown in Tab.\,\ref{table1} and $J_1=2.86$ in the pseudomorphic transition area.}
	\label{fig7}
\end{figure*}

The simulation shown in \fig\ref{fig6}a was performed at 0~K with an applied external field of $B=2$~T. The three different areas of the lattice are indicated by dashed lines. In the initial state a single skyrmion was written in a ferromagnetic background at the position marked by the white cross, its outer rim touching the pseudomorphic Co/Fe area. During the simulation the skyrmion was repelled from the Co/Fe edge and slowly moved further inside the Pd/Fe film. To now investigate the role of magnetic interactions at the Pd/Fe edge, we vary the strength of different parameters in the pseudomorphic Co/Fe area. First, the strength of the exchange interaction is reduced step-wise from $4.86$~meV/atom in \fig\ref{fig6}a to $0.86$~meV/atom in \fig\ref{fig6}f, while DMI is kept constant at the same value as the adjacent Pd/Fe. When $J$ is considerably higher in the pseudomorphic Co/Fe area than in Pd/Fe, the skyrmion is repelled from the boundary during the simulations, as shown in a. In b and c the outer skyrmion rim remains pinned to the Co/Fe edge. This effect can be attributed to the easy-plane anisotropy which is energetically more favorable for the outer skyrmion part than the out-of-plane anisotropy within Pd/Fe. With a smaller $J$, the skyrmion moves further inside the pseudomorphic Co/Fe area, because here a large nearest-neighbor spin angle does not cost as much exchange energy, compared to the Pd/Fe area. The reconstructed Co/Fe area with its relatively high exchange interaction on the other hand serves as a barrier and prevents the skyrmion from entering the Co/Fe completely. When $J$ is sufficiently small the skyrmion strips out along the edge and is fully confined within the Co/Fe transition area, see \fig\ref{fig6}f. We repeat such simulations but now vary the strength of the DMI vector $D=|\mathbf{D}_{i,j}|$. Here the skyrmion is repelled when $D$ becomes very small and strips out into a single spin spiral stripe for a larger $D$. We find that there is a certain parameter range for both $J$ and $D$ in which the skyrmion is stable at the edge, see Supplementary \fig S7. These simulations show that already small parameter changes in the Co/Fe transition area can be the cause of skyrmion pinning at the edge of hcp-stacked Pd/Fe but not at the edge of fcc-stacked Pd/Fe. Furthermore, we find that for the edge-pinning two distinct areas are needed, that together exhibit a Lennard-Jones like potential: A narrow area that is favorable for certain parts of the skyrmion structure, and also a repulsive area, that prevents the skyrmions from fully leaving the Pd/Fe island.

The reconstructed Co/Fe of \fig\ref{fig6} was nearly saturated by the applied magnetic field of $B = 2$~T, making it effectively an out-of-plane magnetized surrounding. We now turn to simulations with smaller magnetic fields, where the in-plane magnetization component of Co/Fe is larger, and investigate its effect on the magnetic state of Pd/Fe near the boundary. In \fig\ref{fig7} simulations were performed both with zero-field and with field-cooling at $B=1.5$~T from a random configuration at 80~K down to 0~K, see methods. To induce different magnetization directions at the boundary of the reconstructed Co/Fe, as observed experimentally, the relative in-plane anisotropy axis $K_x$ was changed as indicated by the white arrows, see \fig\ref{fig7}a. The magnetization of the lowest Co/Fe atomic row was fixed in the same direction, while all other spins were able to relax freely. The simulations at $B=0$~T arrive at very similar spin spiral structures as we have observed experimentally in the magnetic virgin state, cf.\,\fig\ref{fig2}, with spin spiral stripes dominantly parallel to the boundary and zero-field skyrmions in the case of \fig\ref{fig7}e. When the magnetization in Co/Fe points parallel to the boundary, the spin spiral stripes are also dominantly parallel to the edge, and the phase of the spiral, \ie whether the spiral spins are pointing up (red) or down (blue), does not seem to make a difference, see \fig\ref{fig7}b and c. When the magnetization is perpendicular to the edge, see \fig\ref{fig7}d and e, the spiral phase clearly depends on the magnetization direction of the Co/Fe area, which can be attributed to the presence of DMI preferring clockwise rotational sense. 

At $B=1.5$~T the magnetization of the reconstructed Co/Fe area has begun to rotate towards the out-of-plane direction. Yet an in-plane component remains and we can still find differences between different in-plane directions: When the magnetization is parallel to the boundary, see \fig\ref{fig7}f and g, circular skyrmions appear spontaneously at the edge and remain stable during the simulation process. With the magnetization pointing perpendicular to the edge, see \fig\ref{fig7}h and i, we observe the following: First, in \fig\ref{fig7}h a striped-out skyrmion similar to a single spin spiral stripe is stabilized at the edge. Secondly, for the opposite magnetization direction neither compact nor striped-out skyrmions are stable at the edge, as shown in \fig\ref{fig7}i. Here the incorporation of skyrmions at the edge would require an anti-clockwise spin rotation towards the edge, which is unfavorable for the DMI in this system. However, in \fig\ref{fig7}h a clockwise rotation only becomes possible through the inclusion of an edge-skyrmion, see sideview sketch at the bottom of \fig\ref{fig7}a. Therefore skyrmions are repelled in \fig\ref{fig7}i but not in \fig\ref{fig7}h.

\section{Discussion}

Our work demonstrates that by altering the edge properties of a film its magnetic state can be effectively manipulated. Here the ferromagnetic Co/Fe rim changes the propagation direction of the spin spiral in Pd/Fe so that the spiral stripes are parallel to the island edge instead of perpendicular. As shown in the simulations, the in-plane magnetization direction of the rim can also have a small effect on how the spiral connects to the boundary. Furthermore, depending on the size and shape of the Pd/Fe island, the Co/Fe rim enables the formation of zero-field skyrmions and target states in the magnetic virgin state. Judging by the unique spin spiral structures we find in the magnetic virgin state of our system, one might also be able to utilize the edge to stabilize a zoo of different topological spin states like, \eg the recently discussed skyrmion bags~\cite{fosterNP2019,rybakovPRB2019}. 

The Co/Fe rim increases the amount of skyrmions that survive in the remanent state. In fcc-stacked Pd/Fe this happens, because a connection between skyrmions and the boundary is hindered by the ferromagnetic surroundings. In hcp-stacked Pd/Fe the edge-pinned skyrmions strip out along the Co/Fe rim when the field is lowered and hinder further skyrmions from connecting to the edge by repulsive skyrmion-skyrmion interactions. In both cases we anticipate that these additional repulsive forces at the boundary would counteract the SkHE and effectively prevent skyrmion annihilation. Our spin dynamics simulations have shown that there are several material parameters that can be responsible for a repulsion or pinning of skyrmions at the film edge, \eg the edge-pinning observed in hcp Pd/Fe can be caused by either lower exchange interactions, higher DMI, or stronger easy plane anisotropy at the island rim. 

Furthermore, we have observed in experiment and simulations that the magnetization direction at the Co/Fe boundary has an effect on the edge-skyrmions in hcp Pd/Fe. A magnetization parallel to the edge favors the formation of localized skyrmions, while a perpendicular magnetization enables the skyrmions to strip out along the edge even in applied external fields. While we use a self-organized model system to investigate skyrmion physics, we anticipate that these findings can also be applied to top-down approaches to deliberately create structures with different magnetic properties, as it has been shown to be viable in Ref.\,\cite{oharaNL2021} and Ref.\,\cite{jugeNL2021}. Modifying the boundaries of a skyrmion system with tailored materials is a means to set up guard railings, which can improve speed and reliability of potential skyrmion devices.

\begin{figure}[t] 
	\centering	
	\includegraphics[width=0.45\textwidth]{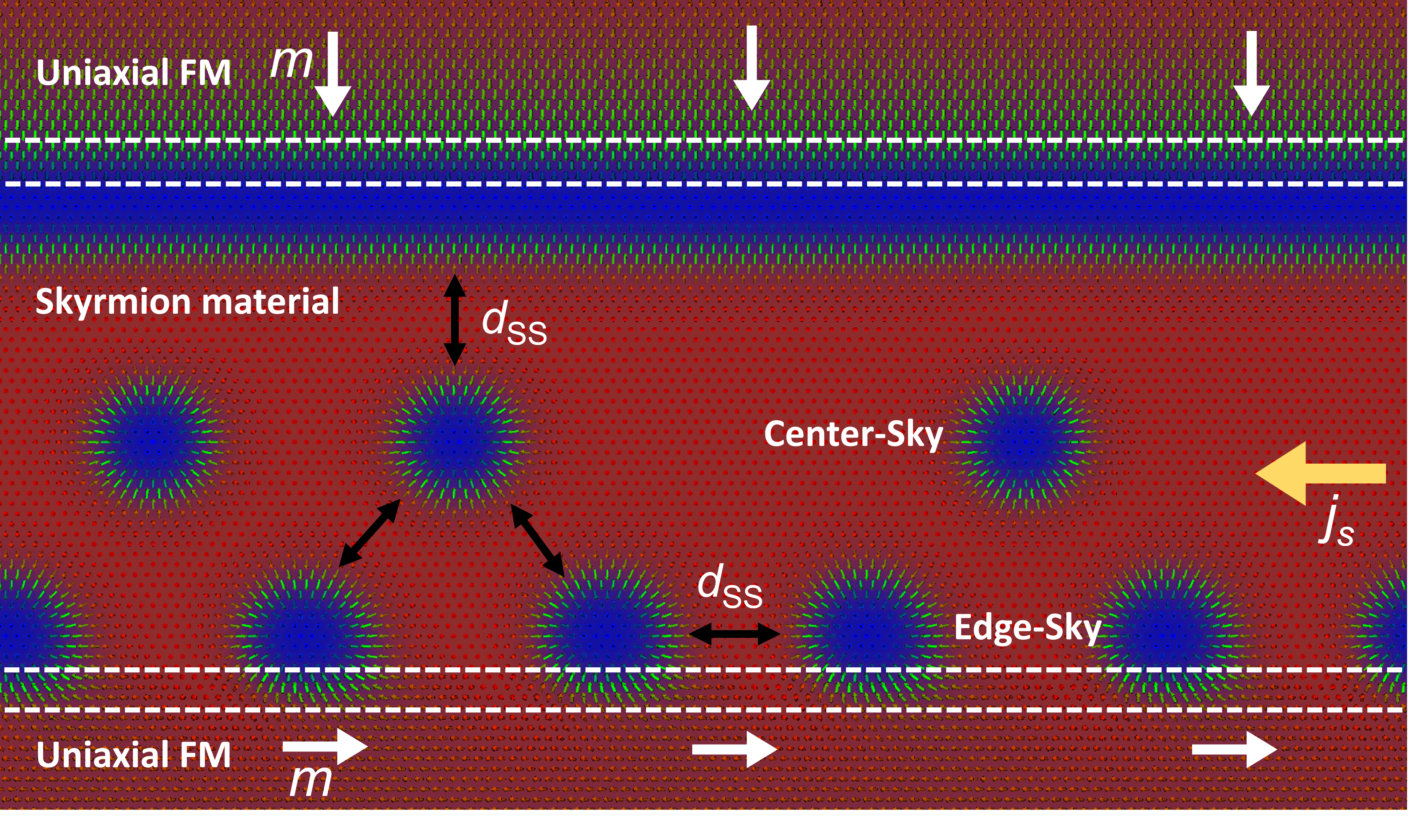}
	\caption{Possible racetrack structure with edge-skyrmions functioning as spacers between skyrmion bits ($B=1.5$~T and $T=0$~K). The material parameters are the same as shown in Tab.\,\ref{table1} with $J=2.86$ in the transition area. At the upper rim, the magnetization of Co/Fe is orthogonal to induce a skyrmion-repelling spiral stripe. }
	\label{fig8}
\end{figure}

Finally we want to propose a possible application for edge-pinned skyrmions: In skyrmion racetrack memory devices the information is stored in the distance between skyrmions~\cite{fertNN2013}. Therefore such devices must rely on equidistant bit positions. Yet there are several factors that can change the distances between skyrmions including thermal drift, noise generated by device operation, as well as attractive or repulsive skyrmion-skyrmion interactions. To overcome these challenges different approaches have been put forward, \eg the use of a two lane racetrack~\cite{mullerIOP2017}, or a periodic arrangement of notches~\cite{fookSR2016}. We propose the use of edge-skyrmions as spacers to divide the track into separated skyrmion-sized bits as shown in \fig\ref{fig8}. In the displayed setup the skyrmion track is confined between two ferromagnetic Co/Fe edges, one with perpendicular and the other with parallel magnetization at the edge. The information is contained in the central row, stored in the absence or presence of a skyrmion along the track. The upper Co/Fe rim repels skyrmions|here realized by inducing a single spin spiral stripe|and forces skyrmions in the center of the track to interact with the edge-skyrmions at the bottom rim. The minimum distance between skyrmions $d_\mathrm{SS}$ leaves enough space for one center-skyrmion to be positioned between two edge-skyrmions and the upper rim. Additionally the repulsive interactions with edge-skyrmions hinders center-skyrmions from hopping from one bit to another. It is anticipated that all skyrmions, the ones that store the information and the ones that are used as bit separators, move coherently under lateral currents. Additionally, small distance variations between bits could be corrected for, because each arrival of an edge-skyrmion near the read/write unit would unambiguously signal the beginning of a new bit.

\section{Methods}

\subsection{Experimental Details}
The samples were prepared and investigated in an ultra-high vacuum system with different chambers for crystal cleaning, metal deposition and STM measurements. The Ir(111) crystal surface was cleaned by cycles of annealing in an oxygen atmosphere of $10^{-6}-10^{-8}$~mbar while slowly ramping the crystal temperature to about $T=1600$~K and subsequent sputtering with Ar-ions. The surface was cured by a final 60~s long electron-beam flash to 1600~K. Approximately $0.7$ atomic layers of Fe were deposited onto the crystal, while the crystal temperature was still elevated ($\approx400$~K), followed by $0.5$ atomic layers of Pd and finally $0.2$ atomic layers of Co. The Co was deposited 30 minutes after the electron-beam flash with the sample now being closer to room temperature.

All STM measurements were performed at 4~K using a Cr bulk tip which can have an arbitrary magnetization axis and varying degrees of spin-polarization~\cite{schlenhoffAPL2010}. Our home-build STM is equipped with a superconducting magnet, that can generate magnetic fields of up to 9~T perpendicular to the sample surface.

\subsection{Simulations}

For the  magnetic parameters of the reconstructed Co/Fe area, we chose to increase the strength of the exchange interaction with respect to Pd/Fe while keeping the DMI contribution at the same value. As an interfacial effect, we anticipate the DMI to mainly act on the Fe layer while the exchange interaction contributes to both layers of Co/Fe. The Co/Fe bilayer exhibits in-plane ferromagnetic domains, magnetized perpendicular to the dislocation lines. We simulate this by adding an in-plane uniaxial anisotropy energy parameter $K_x=0.65$~meV/atom. Furthermore, as shown in Supplementary \fig S6, in a perpendicular field of $B=2$~T the Co/Fe magnetization axis has a large out-of-plane component. We assume that the Co/Fe is more easily magnetized compared to Pd/Fe because its total magnetic moment is larger. However, due to software limitations, we cannot increase the magnetic moment of Co/Fe separately, therefore we instead include an additional out-of-plane anisotropy parameter $K_z$ for the reconstructed Co/Fe and adjust the strength of both anisotropy parameters to reflect our experimental findings and finally arrive at the parameter set shown in Tab.\,\ref{table1}.

Atomistic spin dynamics simulations were performed using the MonteCrystal simulation code which can be found on github~\cite{montecrystal}. The code employs an algorithm that numerically solves the Landau-Lifshitz-Gilbert equation. For all simulations the damping parameter was set to $\alpha=0.5$. The simulated lattice consisted of 80 rows with 100 atomic sites each and periodic boundary conditions in the horizontal direction. The reconstructed Co/Fe was modeled with 10 atomic rows, while the pseudomorphic Co/Fe was 4 atom rows thick. The simulations of \fig\ref{fig6} were carried out for at least 250~ps and were stopped either after they had converged with the skyrmion pinned at the edge or after the skyrmion was repelled. The simulations of \fig\ref{fig7} were started from a random spin configuration and the temperature was reduced in 2~K steps from 80~K to 0~K until the energy reached a stable value.

\section{Data availability}
The data that support the findings of this study are available from the authors upon reasonable request.

\section{Code availability}

The spin dynamics simulations of the manuscript were done with MonteCrystal $3.2.0$, which can be found on github~\cite{montecrystal}. 

\section{Author Contributions}

J.S.\,performed the experiments, J.S.\,and A.K.\,prepared the samples. J.S.\,did the spin dynamics simulations. J.S.\,wrote the manuscript. J.S., E.Y.V., R.W., A.K.\,and K.v.B.\,discussed the results and contributed to the manuscript.

\section{Acknowledgements}
A.K.\,and K.v.B.\,acknowledge financial support from the Deutsche Forschungsgemeinschaft (DFG, German Research Foundation) Grants No.~408119516 and No.~402843438. R.W.\,acknowledges financial support from the ERC (Adv.\ Grant ADMIRE). 

%\bibliographystyle{naturemag}
%\bibliography{CoPdFepaper_1}

\end{document}